\documentclass[11pt,notcite, notref]{article}

\usepackage{graphics,amsmath,amssymb,amsthm,accents}
\usepackage{epic}

\newtheorem{lem}{Lemma}

\def \h#1{\widehat{#1}}
\def \t#1{\widetilde{#1}}
\def \b#1{\overline{#1}}

\def \th#1{\widehat{\widetilde{#1}}}
\def \hb#1{{\widehat{\overline{#1}}}}
\def \bh#1{{\widehat{\overline{#1}}}}

\def \tb#1{\widetilde{\overline{#1}}}
\def \bt#1{\widetilde{\overline{#1}}}
\def \dh#1{\underaccent{\hat}{#1}}

\def \dt#1{\underaccent{\tilde}{#1}}
\def \dth#1{\underaccent{\tilde}{\underaccent{\hat}{#1}}}

\numberwithin{equation}{section}
\newtheorem{prop}{Proposition}

\title{Multisoliton solutions to the lattice Boussinesq equation}
\author{Jarmo Hietarinta$^1$\footnote{E-mail: jarmo.hietarinta@utu.fi}
  ~ and ~ Da-jun Zhang$^2$\footnote{E-mail: djzhang@staff.shu.edu.cn }
  \\
  {\small\it $^1$Department of Physics and Astronomy,
    University of Turku, FIN-20014 Turku, Finland} \\
  {\small\it $^2$Department of Mathematics, Shanghai University,
    Shanghai 200444, P.R. China}}

\date{\today}
\begin{document}

\maketitle

\begin{abstract}
  The lattice Boussinesq equation (BSQ) is a three-component
  difference-difference equation defined on an elementary square of
  the 2D lattice, having 3D consistency. We write the equations in the
  Hirota bilinear form and construct their multisoliton solutions in
  terms of Casoratians, following the methodology in our previous
  papers. In the construction it turns out that instead of the usual
  discretization of the exponential as $[(a+k)/(a-k)]^n$ we need two
  different terms $[(a-\omega k)/(a-k)]^n$ and $[(a-\omega^2
  k)/(a-k)]^n$, where $\omega$ is a cubic root of unity $\neq 1$.
\end{abstract}

\section{Introduction}
Among the integrable 2D difference equations one important class
consists of those equations that are defined on an elementary
quadrilateral and are multidimensional consistent. Here
multidimensional consistency means that one can add a third dimension
and extend the definition in a natural way from a quadrilateral to a
cube, and that on this cube the maps are consistent.

Most of the maps in this class are included in the ABS classification
\cite{ABS-CMP-2002}, (which is complete within the assumptions of
symmetry and the tetrahedron property). These lattice maps have one
component for each lattice site, but there are also 3D-consistent
{\em multicomponent} maps related to the Boussinesq (BSQ) equation
\cite{NPCQ-Inv-1992, N-rew-BobSei-1999, Walker, TN-Bous-2005}.

In this paper we will derive multisoliton solutions to the lattice BSQ
equation defined on the elementary square by the equations
\cite{TN-Bous-2005}
\begin{subequations}\label{DB}
\begin{align}
B1 & \equiv\t w - u\t u+v=0,\label{DB-a}\\
B2 & \equiv \h w - u\h{u}+v=0,\label{DB-b}\\
B3 &\equiv w - u\th{u}+\th{v}-\frac{p-q}{\h{u}-\t{u}}=0,\label{DB-c}
\end{align}
\end{subequations}
where we have used the standard shorthand notation, e.g., $\t
u=u_{n+1,m},\, \h v=v_{n,m+1}$, and where $p$ and $q$ are
parameters in the $n$ and $m$  directions, respectively.
From equations \eqref{DB-a},\eqref{DB-b} one finds, after taking the hat
and tilde shifts, respectively
\begin{subequations}
\begin{align}
\h{\t u}&=\frac{\t v-\h v}{\t u-\h u},\\
\h{\t w}&=\frac{\t u\h v-\h u\t v}{\t u-\h u}.
\end{align}
\label{DB-s}
\end{subequations}
Equations \eqref{DB},\eqref{DB-s} are consistent on the elementary
square.

Three-dimensional consistency means that we can add a third direction,
with parameter $r$. The new equations will be
\begin{subequations}
\begin{align}
\b w - u\b u+v=&0,\label{DB3-a}\\
w - u\bh{u}+\bh{v}-\frac{r-q}{\h{u}-\b{u}}=&0,\label{DB3-b}\\
w - u\bt{u}+\bt{v}-\frac{r-p}{\t{u}-\b{u}}=&0,\label{DB3-c}
\end{align}
\label{DB3}
\end{subequations}
where we have denoted the shift in the third direction by a bar. To
this we should add the bar-tilde and bar-hat versions of \eqref{DB-s},
which can be derived from \eqref{DB3-a}.

\begin{figure}[h]
\setlength{\unitlength}{0.0004in}
\hspace{2cm}
\hspace{4cm}
\begin{picture}(3482,3700)(0,-500)
\put(450,1883){\circle*{150}}
\put(-100,1883){\makebox(0,0)[lb]{${\bt F}$}}
\put(1275,2708){\circle*{150}}
\put(825,2708){\makebox(0,0)[lb]{$\b F$}}
\put(3075,2708){\circle*{150}}
\put(3375,2633){\makebox(0,0)[lb]{${\bh F}$}}
\put(2250,83){\circle*{150}}
\put(2650,8){\makebox(0,0)[lb]{${\th F}$}}
\put(1275,908){\circle*{150}}
\put(1275,908){\circle*{150}}
\put(825,908){\makebox(0,0)[lb]{$F$}}
\put(2250,1883){\circle*{150}}
\put(1950,2000){\makebox(0,0)[lb]{${\th{\b F}}$}}
\put(450,83){\circle*{150}}
\put(0,8){\makebox(0,0)[lb]{$\t F$}}
\put(3075,908){\circle*{150}}
\put(3300,833){\makebox(0,0)[lb]{$\h F$}}
\drawline(1275,2708)(3075,2708)
\drawline(1275,2708)(450,1883)
\drawline(450,1883)(450,83)
\drawline(3075,2708)(2250,1883)
\drawline(450,1883)(2250,1883)
\drawline(3075,2633)(3075,908)
\dashline{60.000}(1275,908)(450,83)
\dashline{60.000}(1275,908)(3075,908)
\drawline(2250,1883)(2250,83)
\drawline(450,83)(2250,83)
\drawline(3075,908)(2250,83)
\dashline{60.000}(1275,2633)(1275,908)
\end{picture}
\caption{The multi-dimensional
  consistency cube.\label{F:1}}
\end{figure}
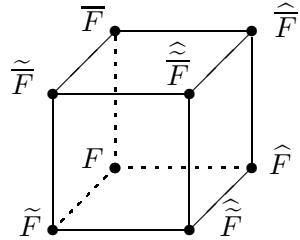

Consider now the cube of Figure \ref{F:1}, where $F$ stands for the
three components $(u,v,w)$.
The initial values are given at $F,\,\t F,\,\h F,\,\b F$, but due to
the linear equations \eqref{DB-a},\eqref{DB-b},\eqref{DB3-a} we only need
to give $(u,v,w)$, $(\t u,\t v)$, $(\h u,\h v)$, $(\b u,\b v)$. Given
these values we can compute $(\th u, \th v, \th w)$, $(\bh u, \bh v,
\bh w)$ $(\bt u, \bt v, \bt w)$ using \eqref{DB-s} and its tilde-bar
and hat-bar versions, along with \eqref{DB-c},\eqref{DB3-b},\eqref{DB3-c}.
After this there are three different ways to compute the remaining
$(\b{\h{\t u}},\b{\h{\t v}},\b{\h{\t w}})$ but they all give the same
values and the system is therefore 3D-consistent.

In the following construction we make active use the 3D-view with
different interpretations for the bar shift. In Section \ref{sec2} we
construct first the background solution, then the one-soliton
solution, and then in Section \ref{sec3} we derive the Hirota bilinear
form and propose and prove the formulae for $N$-soliton solutions. The
procedure is in principle similar to the one used in
\cite{Sol-Q4-2007,Sol-Q3-2008,HZ-PartII}, but the multicomponent
nature induces some new features.

\section{The background and one-soliton solutions\label{sec2}}
The fixed-point idea proposed in \cite{Sol-Q4-2007} for the
construction of the background solution (0SS) is that the background
or ``seed'' solution should be a fixed point with respect to a shift
in the third direction, i.e., $\b u=u,\b v=v,\b w=w$. The relevant
equations (the equations on the sides of the cube) are then obtained
from \eqref{DB3} as
\begin{equation}
\left\{\begin{array}{l}
w = u^2-v, \\
w = u\t{u}-\t{v}+\frac{p-r}{{u}-\t{u}},\\
w = u\h{u}-\h{v}+\frac{r-q}{\h{u}- {u}},
\end{array}\right.
\end{equation}
where $r$ is the parameter in the bar-direction and now plays the role
of a parameter of the background solution. The equations are easy to
solve and we find the 0SS
\begin{subequations}
\begin{align}
u_0 &= an+bm+c_1,\label{u0}\\
v_0 &= \tfrac{1}{2}u^2_0+\tfrac{1}{2}(a^2n+b^2m+c_2)+c_3,\label{v0}\\
w_0 &=\tfrac{1}{2}u^2_0-\tfrac{1}{2}(a^2n+b^2m+c_2)-c_3,\label{w0}
\end{align}
\label{0SS}
\end{subequations}
where $a,b$ are related to $p,q$ by
\begin{equation}
a^3=r-p,~~b^3=r-q,
\label{ab-pq}
\end{equation}
and $c_1,c_2,c_3$ are arbitrary constants.

The one-soliton solution (1SS) is constructed with the same idea of
using the 3D cube, but now $u,v,w$ correspond to the background
solution and $\b u,\b v,\b w$ to the 1SS. In more detail, we have the
side equations
\begin{subequations}\label{1ss-side}
  \begin{equation}
    \label{1ss-w}
    \b w = u\b u-v,
  \end{equation}
together with
\begin{equation}
\left\{\begin{array}{l}
\bt u = \frac{\t{v}-\b{v}}{\t{u}-\b{u}}, \\
\tb{v} = u\tb{u}-w+\frac{k^3-a^3}{\b{u}-\t{u}},
\end{array}\right.
\quad
\left\{\begin{array}{l}
\hb{u} = \frac{\b{v}-\h{v}}{\b{u}-\h{u}}, \\
\hb{v}= u\hb{u}-w+\frac{b^3-k^3}{\h{u}- \b{u}}.
\end{array}\right.
\label{BT}
\end{equation}
\end{subequations}
These act as a B\"acklund transformation (BT) with $k$ as the BT
parameter related to the bar direction.

In order to solve equations \eqref{1ss-side} we take
$(u,v,w)=(u_0,v_0,w_0)$ where the background solution is defined in
\eqref{0SS}. The 1SS is written in the form
\begin{equation}
(\b{u},\b{v},\b{w})=(\b{u}_0+x,\b{v}_0+y,\b{w}_0+z),
\label{bar-1SS}
\end{equation}
where $(\b{u}_0,\b{v}_0,\b{w}_0)$ is the bar-shifted $(u_0,v_0,w_0)$, i.e.,
\begin{subequations}
\begin{align}
  \b u_0 &= an+bm+k+c_1,\label{bu0}\\
  \b v_0 &= \tfrac{1}{2}{\b u}^2_0+\tfrac{1}{2}(a^2n+b^2m+k^2+c_2)+
  c_3,\label{bv0}\\
  \b w_0 &=\tfrac{1}{2}{\b
    u}^2_0-\tfrac{1}{2}(a^2n+b^2m+k^2+c_2)-c_3.\label{bw0}
\end{align}
\label{bar-0SS}
\end{subequations}

With these definitions we find from \eqref{1ss-w} that $z=u_0x$. Thus
we only need to solve for $x,y$, for which we have from \eqref{BT}
\begin{equation}
  \label{BT-eqs}
\left\{\begin{array}{rcl}
  \t x&=&\frac{-\b{\t u}_0\, x+y}{x-a+k},\\
  \h x&=&\frac{-\b{\h u}_0\, x+y}{x-b+k},
\end{array}\right.\quad
\left\{\begin{array}{rcl}
  \t y&=&\frac{-(\b{\t v}_0+w_0)\,x+u_0\,y}{x-a+k},\\
  \h y&=&\frac{-(\b{\h v}_0+w_0)\,x+u_0\,y}{x-b+k},\\
\end{array}\right.\quad
\end{equation}
This system can be linearized by taking
$(x,y)=(\frac{G}{F},\frac{H}{F})$, the result is
\begin{subequations}
\begin{equation}
\t\Psi=N\Psi,\quad\h\Psi=M\Psi, \label{Psi}
\end{equation}
where
\begin{equation}
\Psi=\begin{pmatrix}G\\ H\\ F\end{pmatrix},~~
N=\begin{pmatrix}
                     \tb u_0 &-1 &0\\
                     \tb v_0+w_0 &-u_0 &0\\
                     -1 & 0 & a-k
        \end{pmatrix},~~
M=\begin{pmatrix}
                     \hb  u_0 &-1 &0\\
                     \hb v_0+w_0 &-u_0 &0\\
                     -1 & 0 & b-k
        \end{pmatrix}.
\end{equation}
\end{subequations}
The matrices $N,M$ satisfy the integrability condition $\h N M=\t M
N$.

In order to construct the solutions it is useful to note that if we
define
\begin{equation}
  \label{dnm}
  Q(n,m)=\begin{pmatrix}
u_0(n,m)-\omega k & -1 & 0 \\
u_0(n,m)-\omega^2 k & -1 & 0 \\
(-u_0(n,m)+ k)/(3k^2) & 1/(3k^2) & 1
\end{pmatrix},
\end{equation}
then
\begin{equation}
  \label{dnm2}
  N=Q(n+1,m)^{-1}\,D(a)\,Q(n,m),\quad \text{where}\quad
D(a)=\begin{pmatrix}
a-\omega k & 0 & 0 \\
0 & a-\omega^2 k & 0 \\
0 & 0 & a-k
\end{pmatrix},
\end{equation}
and similarly $ M=Q(n,m+1)^{-1}\,D(b)\,Q(n,m)$. Here $\omega$ is a
cubic root of unity, $\omega\neq 1$, i.e, $\omega^2+\omega+1=0$.
Using these  we can straightforwardly construct the solution as
\begin{equation}
  \label{sol-1ss}
  \Psi(n,m)=Q(n,m)^{-1}D(a)^nD(b)^mQ(0,0)\Psi(0,0),
\end{equation}
from which we find
\begin{subequations}
\begin{eqnarray}
  \label{1ss}
  G&=&k(\omega-1)\left[\rho^0_1(a-\omega k)^n(b-\omega k)^m
  -\omega^2\rho^0_2(a-\omega^2 k)^n(b-\omega^2 k)^m\right],\\
  H&=&u_0\,g+k^2(\omega-1)\left[-\omega^2\rho^0_1(a-\omega k)^n(b-\omega k)^m
  +\rho^0_2(a-\omega^2 k)^n(b-\omega^2 k)^m\right],\\
  F&=&\rho^0_0(a-k)^n(b-k)^m
+\rho^0_1(a-\omega k)^n(b-\omega k)^m+\rho^0_2(a-\omega^2
k)^n(b-\omega^2 k)^m,
\end{eqnarray}
\end{subequations}
where we have introduced new constants $\rho_\nu^0$ in place of
$G_{00},H_{00},F_{00}$. Using \eqref{bar-1SS},\eqref{bar-0SS} we can
recover the 1SS as
\begin{subequations}
  \label{1ss-fin}
\begin{align}
&u^{1SS}=u_0+k\frac{1+\omega\rho_1+ \omega^2\rho_2}{1+\rho_1+\rho_2},\\
&v^{1SS}=v_0+u_0\, k\frac{1+\omega\rho_1+ \omega^2\rho_2}{1+\rho_1+\rho_2}
+k^2\,\frac{1+\omega^2\rho_1+ \omega\rho_2}{1+\rho_1+\rho_2},\\
&w^{1SS}=w_0+u_0\, k\frac{1+\omega\rho_1+ \omega^2\rho_2}{1+\rho_1+\rho_2}.
\end{align}
\end{subequations}
Here $u_0,v_0,w_0$ were defined in \eqref{0SS} and
\begin{equation}\label{rho}
\rho_\nu(n,m;k)=\frac{(a-\omega^\nu k)^n}{(a- k)^n}
\frac{(b-\omega^\nu k)^m}{(b- k)^m}\frac{\rho_\nu^0}{\rho_0^0},\quad \nu=1,2,
\end{equation}
where $k$ is the soliton parameter. 

\section{Bilinearization, Casoratians and $N$-soliton
  solutions\label{sec3}}
\subsection{The main result}
The 1SS \eqref{1ss-fin} is not quite sufficient for guessing the
general structure for NSS, but after considering also 2SS (in the case
$\rho_2^0=0$) we arrived to a solution in terms of Casoratians,
constructed as follows: Given the multi-indexed function
\begin{equation}
  \psi_j(n,m,l)=\sum^{3}_{s=1}\varrho^{(0)}_{j,s}
  (\delta-\omega^s k_j)^l (a-\omega^s k_j)^n (b-\omega^s k_j)^m,
\label{psi-gen}
\end{equation}
we define the column vector
\begin{equation}\label{C-entry-vec}
\psi(n,m,l)=(\psi_1(n,m,l),\psi_2(n,m,l),\cdots,\psi_{N}(n,m,l))^T,
\end{equation}
and then the generic $N\times N$ Casorati matrix by combining
columns with different shifts $l_i$. The generic Casoratian is then the
determinant
\begin{equation}
  \label{eq:C-gen}
  C_{n,m}(\psi;\{l_i\})=
  |\psi(n,m,l_1),\psi(n,m,l_2),\cdots,\psi(n,m,l_N)|.
\end{equation}
To describe such Casoratians we use the  shorthand notation
\cite{Freeman-Nimmo-KP} in which only the shifts are
given. Furthermore for consecutive sequences we use
$\h{M}\equiv 0,1,\dots,M$ (this cannot be confused with the use of hat
for shifts).

\begin{prop}
Multisoliton solutions to \eqref{DB} are given by
\begin{equation}
u=u_0-\frac{g}{f},~~~v=v_0-u_0 \frac{g}{f}+\frac{h}{f},~~~
w=w_0-u_0 \frac{g}{f}+\frac{s}{f},
\label{trans}
\end{equation}
where $u_0,v_0,w_0$ are given in \eqref{0SS} and the functions
$f,g,h,s$ are given in terms of Casoratians\footnote{
If $N=1$ then $f=|0|,g=|1|,h=|2|, s=0$;
and if $N=2$ then $f=|0,1|,g=|0,2|,h=|0,3|, s=|1,2|$.}
\begin{equation}
\label{casdef}
  f=|\h{N-1}|,\,  g=|\h{N-2},N|,
\,  h=|\h{N-2},N+1|,\, s=|\h{N-3},N-1,N|,
\end{equation}
composed of $\psi$ given in \eqref{psi-gen} with $\delta=0$.
\end{prop}

Here the size on the matrix $N$ indicates the number of solitons and the
set $\{k_i\}_{i=1}^N$ provides the ``velocity'' parameters of the
solitons, while the parameters $\varrho_{j,s}^{(0)}$ are
related to the locations of the solitons (by gauge invariance only
their ratio is significant).

In order to prove the above Proposition, we note that using
\eqref{trans} as a dependent variable transformation we can
bilinearize \eqref{DB} as
\begin{subequations}
\begin{align}
  \mathcal{B}_1&=\t{f}(h+ag)-\t{g}(g+af)+f\t{s}=0, \label{bil-DB-a}\\
  \mathcal{B}_2&=\h{f}(h+bg)-\h{g}(g+bf)+f\h{s}=0,\label{bil-DB-b}\\
  \mathcal{B}_3&=\t{f}\h{g}-\h{f}\t{g}-(a-b)(\t{f}\h{f}-f\th{f})=0,
\label{bil-DB-c}\\
  \mathcal{B}_4&=(a^2+ab+b^2)(f\th{f}-\t{f}\h{f})+
(a+b)(f\th{g}-\th{f}g)+\th{f}s+f\th{h}-g\th{g}=0.\label{bil-DB-d}
\end{align}
\label{bil-DB-I}
\end{subequations}
In fact, the lattice BSQ equation \eqref{DB} can be written as
\begin{equation}
  B1 =\frac{\mathcal{B}_1}{f\t{f}}\quad B2  =\frac{\mathcal{B}_2}{f\h{f}}\quad
  (\h u-\t u)B3
  =\frac{\mathcal{B}_3\mathcal{B}_4-(a-b)f\th{f}\mathcal{B}_4
+(a^2+ab+b^2)\t{f}\h{f}\mathcal{B}_3}{f\t{f}\h{f}\th{f}}.
\end{equation}
Thus we need to prove that the Casoratians \eqref{casdef} solve the
bilinear equations \eqref{bil-DB-I}. This is given in Appendix
\ref{A} for the generic values of $\delta$, although only $\delta=0$
is used for the lattice BSQ. The Casoratian proof suggested the $\delta\neq0$
generalization along with some others, they are discussed next.

\subsection{The role of $\delta$}
The solution to the lattice BSQ are obtained with $\delta=0$ in the
matrix entry \eqref{psi-gen} but the generalization $\delta\neq 0$ is
natural and we may ask about its meaning. It is important to note that
the proof given in the Appendix can be carried out using only the
following assumptions on the column vectors
\begin{subequations}
\label{cas-cond}
\begin{eqnarray}
(a-\delta)  \dt \psi &=&\psi -\dt{\b \psi},
\label{rela-I-a-1}\\
(b-\delta)  \dh \psi&=&\psi-\dh{\b \psi},
\label{rela-I-a-2}\\
\gamma \psi &=&\b{\b{\b{\psi}}} -3\delta\b{\b{\psi}} +3\delta^2\b{\psi},
\label{rela-I-a-3}
\end{eqnarray}\end{subequations}
where $\gamma$ is a diagonal matrix and the bar-shift of $\psi$ is
defined by $\b\psi(n,m,l)=\psi(n,m,l+1)$. From these assumptions one
can derive the bilinear equations
\begin{subequations}
\label{bil-DB-II}
\begin{align}
  \mathcal{B}^{\delta}_1&=\t{f}[h+(a-\delta)g]-\t{g}[g+(a-\delta)f]+f\t{s}=0,
\label{bil-DB-a2}\\
  \mathcal{B}^{\delta}_2&=\h{f}[h+(b-\delta)g]-\h{g}[g+(b-\delta)f]+f\h{s}=0,
\label{bil-DB-b2}\\
  \mathcal{B}^{\delta}_3&=\t{f}\h{g}-\h{f}\t{g}+(a-b)(\t{f}\h{f}-f\th{f})=0,
  \label{bil-DB-c2}\\
  \mathcal{B}^{\delta}_4&=(a^2+ab+b^2)(f\th{f}-\t{f}\h{f})+
  (a+b+\delta)(\th{f}g-f\th{g})+\th{f}s+f\th{h}-g\th{g}=0.\label{bil-DB-d2}
\end{align}
\end{subequations}

We can now reverse the dependent variable transformation \eqref{trans}
and construct from \eqref{bil-DB-II} a generalized lattice BSQ
equation
\begin{subequations}
\begin{align}
\t w &= u\t u-v+\delta(\t{u}-u-a),\label{DB-a2}\\
\h w &= u\h{u}-v+\delta(\h{u}-u-b),\label{DB-b2}\\
w &= u\th{u}-\th{v}+\frac{-a^3+b^3}{\h{u}-\t{u}}-\delta(\th{u}-u-a-b).
\label{DB-c2}
\end{align}
\label{DB-II}
\end{subequations}
Obviously, if we take $\delta=0$, the above equations reduce to those
of lattice BSQ.

Now considering the Casoratian forms of $f,g,h,s$ one can easily show
that
\begin{eqnarray*}
  f(\delta)&=&f(0),\\
  g(\delta)&=&g(0)+N\,\delta\,f(0),\\
  h(\delta)&=&h(0)+(N+1)\,\delta\,g(0)+N(N+1)/2\,\delta^2\,f(0),\\
  s(\delta)&=&s(0)+(N-1)\,\delta\,g(0)+N(N-1)/2\,\delta^2\,f(0),
\end{eqnarray*}
where $N$ is the dimension of the matrix. From this result, which was
derived from a particular form of the solution, we are led to the
following: If we denote the $\delta$-dependent functions in
\eqref{DB-II} by $u',v',w'$ then the transformation
\begin{equation}
u'=u,\quad v'=v-\delta\, (u-u_0),\quad w'=w+\delta\, (u-u_0),
\label{gauge}
\end{equation}
converts \eqref{DB-II} into \eqref{DB}. Thus the effect of introducing
$\delta$ in \eqref{psi-gen} can be undone by the ``gauge'' transformation
\eqref{gauge}.

\subsection{Toeplitz generalization}
Another generalization in the construction of the solution is obtained
if we replace \eqref{rela-I-a-3} with
\begin{equation}
  \label{rela-Gamma}
  \Gamma \psi =\b{\b{\b{\psi}}} -3\delta\b{\b{\psi}} +3\delta^2\b{\psi},
\end{equation}
where $\Gamma$ is some $N\times N$ matrix. Under this generalization
proof of the bilinear equations \eqref{bil-DB-II} proceeds as before,
except for some details described in Appendix \ref{G-proof}.

First note that since the solutions are given in terms of determinants
any matrix similar to $\Gamma$ yields same solution as $\Gamma$.
Thus it is sufficient to consider only different canonical forms of
$\Gamma$.

If $\Gamma$ is a diagonal matrix
\begin{subequations}\label{gamma-dia}
\begin{equation}
\Gamma=\mathrm{Diag}(\gamma_1,\gamma_2,\cdots,\gamma_{N})
\end{equation}
where
\begin{equation}
\gamma_j=\delta^3- k_j^3,
\label{gamma-j}
\end{equation}
\end{subequations}
with distinct $\{k_j\}$, then the entries in the Casoratian can be as
in \eqref{psi-gen}.  Since $\omega^2=\omega^*$ (where $*$ stands for
complex conjugate), the condition for a real solution (coming from
real $\psi_j$) is
\begin{equation}
  \varrho^{(0)}_{j,2}=\varrho^{(0)*}_{j,1},\quad
\mathrm{with}\quad k_j,\varrho^{(0)}_{j,3}\in \mathbb{R}.
\label{real-cond}
\end{equation}

Next suppose that $\Gamma$ is a lower triangular matrix defined as
\begin{equation}
\Gamma=\Gamma_N(k_1)=(\gamma_{s,l}(k_1))_{N\times N},~~~
\gamma_{s,l}(k_1)=\left\{\begin{array}{ll}
                    \frac{1}{(s-l)!}\partial^{s-l}_{k_1}\gamma_1,& s\geq l,\\
                    0,                                             & s<l,
                    \end{array}\right.
\label{gamma-jor}
\end{equation}
where $\gamma_1$ is defined by \eqref{gamma-j}. In this case, the
generic entry vector $\psi$ can be taken as
\begin{subequations}
\begin{equation}
\psi=\sum^{3}_{s=1}\mathcal{A}_s\mathcal{Q}_s(k_1),
\label{psi-jor}
\end{equation}
where
\begin{equation}
\begin{array}{l}
\mathcal{Q}_s(k_1)=(Q_{s,0}(k_1),Q_{s,1}(k_1),\cdots, Q_{s,N-1}(k_1))^T,\\
Q_{s,j}(k_1)=\frac{1}{j!}\partial^{j}_{k_1}\left[\varrho^{(0)}_{1,s}
  (\delta-\omega^s k_1)^l (a-\omega^s k_1)^n (b-\omega^s k_1)^m\right],
\end{array}
\label{Q-kj}
\end{equation}
\end{subequations}
and $\{\mathcal{A}_s\}$ are arbitrary $N$th-order lower triangular
Toeplitz matrices defined by
\begin{equation*}
\mathcal{A}_{s}=\left(\begin{array}{cccccc}
a_{s,0} & 0    & 0   & \cdots & 0   & 0 \\
a_{s,1} & a_{s,0}  & 0   & \cdots & 0   & 0 \\
a_{s,2} & a_{s,1}  & a_{s,0} & \cdots & 0   & 0 \\
\cdots &\cdots &\cdots &\cdots &\cdots &\cdots \\
a_{s,N-1} & a_{s,N-2} & a_{s,N-3}  & \cdots &  a_{s,1}   & a_{s,0}
\end{array}\right)_{N\times N},
\quad a_{s,j}\in \mathbb{C}.
\end{equation*}
To get a real solution, in addition to the condition \eqref{real-cond}
we also need $\mathcal{A}_2=\mathcal{A}_1^*$ and $\mathcal{A}_3$ being
real.  Reduction to the solution of the lattice BSQ \eqref{DB} are
obtained by taking $\delta=0$ in \eqref{Q-kj}.

Note that in this case there is only one $k_1$, so solutions
corresponding to such a $\Gamma_N(k_1)$ are kind of limit solutions.
In fact, let us start from the Casoratians with $\psi_j$ defined in
\eqref{psi-gen} and set $\varrho^{(0)}_{j,s}=\varrho^{(0)}_{1,s}$ for
$j=2,3,\cdots, N$.  First we replace the Casoratians $f,g,h,s$ in the
bilinear equations \eqref{bil-DB-II} by $f/K,g/K,h/K,s/K$ with
$K=\prod^{N}_{j=2}\frac{(k_j-k_1)^{j-1}}{(j-1)!}$.  Then, using
L'Hospital rule we can take the limits $k_j\to k_1$ from $j=2$ to
$j=N$ step by step, and finally reach the bilinear equations
\eqref{bil-DB-II} solved by Casoratians $f,g,h,s$ with entry vector
$\psi=\sum^{3}_{s=1}\mathcal{Q}_s(k_1)$ where $\mathcal{Q}_s(k_1)$ is
defined as \eqref{Q-kj}.  The Toeplitz matrix $\mathcal{A}_s$ in
\eqref{psi-jor} can be obtained by a suitable redefinition of
$\varrho^{(0)}_{1,s}$ as a function sufficiently differentiable
w.r.t $k_1$ and $\varrho^{(0)}_{j,s}=\varrho^{(0)}_{1,s}(k_j)$
for $j=2,3,\cdots, N$.

From the above solution we can also derive rational solutions, by
taking a particular choice of parameters in the limit $k_1\to 0$.

\section{Conclusions}
In this paper we have bilinearized the lattice Boussinesq equation
\eqref{DB} and constructed its multi-soliton solutions in terms of
Casoratians. The method, which relies heavily on 3D-consistency, is
similar to the one used in \cite{Sol-Q3-2008,HZ-PartII}, with the main
difference that the analogue of the exponential factor now contains
cubic roots of unity, see \eqref{rho}, allowing two different
$\rho$-terms in the bilinear construction. Cubic roots of unity also
enter in the continuum case, because the continuum Boussinesq equation
is obtained as a three-reduction from the KP equation.

After this work was completed we were informed about reference
\cite{MKO} where the authors bilinearized the lattice BSQ in its
9-point 1-component form using singularity confinement and then
constructed its solutions in Casoratian form. Their final result is
similar to the present except that it only contained two terms in the
Casoratian entries (cf. \eqref{psi-gen}).

\section*{Acknowledgments}
The authors thank F. Nijhoff and K. Kajiwara for discussions and for
providing the references \cite{Walker} and \cite{MKO}, respectively.
This work was finished at the Isaac Newton Institute for Mathematical
Sciences and the authors thank the Institute for hospitality.  One of
the authors (D-j Zh) was supported by the National Natural Science
Foundation of China (10671121), Shanghai Leading Academic Discipline
Project (No.J50101) and Shanghai Magnolia Grant (2008B048).

\vskip 1cm

\begin{appendix}
\section{Proof of Casoratian solutions\label{A}}
Usually in Casoratian proof the bilinear equation is reduced to a
Laplace expansion of a $2N\times 2N$ determinant that can be seen to
be identically zero. The expansion is described as:
\begin{lem}\cite{Freeman-Nimmo-KP}
\label{L:lap}
Suppose that $\mathbf{B}$ is an $N\times(N-2)$ matrix and
$\mathbf{a},~ \mathbf{b},~ \mathbf{c},~ \mathbf{d}$ are $N$'th-order
column vectors, then
\begin{equation}
|\mathbf{B},\mathbf{a}, \mathbf{b}||\mathbf{B}, \mathbf{c},\mathbf{d}|
-|\mathbf{B}, \mathbf{a}, \mathbf{c}||\mathbf{B},\mathbf{b},\mathbf{d}|
+|\mathbf{B},\mathbf{a},\mathbf{d}||\mathbf{B},\mathbf{b},\mathbf{c}|=0.
\label{laplace}
\end{equation}
\end{lem}
This Lemma is also used to generate Casoratian equalities by means of
which one can simplify Casoratian proofs.

\subsection{Formulae for Casoratians\label{A:1}}
In order to use the above Lemma we need various formulae for the
shifts of the Casoratians $f,g,h,s$ of \eqref{casdef} with entries
\eqref{psi-gen}, as given below.  These formulae can be derived using
\eqref{rela-I-a-1},\eqref{rela-I-a-2} in the same way as in
\cite{HZ-PartII}.  For convenience we introduce an
up-shift operator $E^{\nu}$ by
\[
E^{1}\psi\equiv \t\psi,\quad E^{2}\psi\equiv \h\psi,\quad
E^{3}\psi\equiv \b\psi.
\]
Down shifts are denoted by $E_{\nu}$, $\nu=1,2,3$, respectively.

The basic shift formulae are
\begin{subequations}
\label{Formula-I}
\begin{align}
  -(\alpha_\mu-\delta)^{N-2} E_{\mu} f &
  =|\h{N-2},E_{\mu} \psi(N-2)|,\label{f-mu}\\
  -(\alpha_\mu-\delta)^{N-2} E_{\mu} [g+(\alpha_\mu\!-\delta)f]
  &= |\h{N-3},N-1,E_{\mu} \psi(N-2)|,\label{fg-mu}\\
-(\alpha_\mu-\delta)^{N-2} E_{\mu} [h+(\alpha_\mu-\delta)g]
  &= |\h{N-3},N,E_{\mu} \psi(N-2)|,\label{gh-mu}\\
  (a-b)(a-\delta)^{N-2}(b-\delta)^{N-2}\dth{f}
  & = |\h{N-3},\dh \psi(N-2),\dt \psi(N-2)|,\label{f-th}
\end{align}
\begin{align}
  (a-b)&(a-\delta)^{N-2}(b-\delta)^{N-2}\Bigl[\,\dth{g}
  +(a+b-2\delta)\dth{f}\Bigr]\nonumber\\
  =& -(a-\delta)^{N-2}\dt{f}+(b-\delta)^{N-2}\dh{f}
  +|\h{N-4},N-2,\dh \psi(N-2),\dt \psi(N-2)|,\label{g-th}\\
  (a-b)&(a-\delta)^{N-2}(b-\delta)^{N-2}\Bigl[\dth{s}+
  (a+b-2\delta)\dth{g}+[(a-\delta)^2+(a-\delta)(b-\delta)
  +(b-\delta)^2]\dth{f}\Bigr]\nonumber\\
  =&
  -(b-\delta)(a-\delta)^{N-2}\dt{f}+(a-\delta)(b-\delta)^{N-2}\dh{f}
  +|\h{N-5},N-3,N-2,\dh \psi(N-2),\dt \psi(N-2)|.\label{s-th}
\end{align}
\end{subequations}
where $\mu=1,2$, and $\alpha_1=a,\alpha_2=b$.

In order to apply Lemma \ref{L:lap} in the proof of the main result we
need some further equalities (also derived using Lemma \ref{L:lap}):
\begin{subequations}
\label{id-I}
\begin{align}
  f|\h{N-5},N-3,&N-2,\dh \psi(N-2),\dt \psi(N-2)|\nonumber\\
  =&  -(b-\delta)^{N-2}\dh{f}|\h{N-5},N-3,N-2,N-1,\dt \psi(N-2)|\nonumber\\
  & + (a-\delta)^{N-2}\dt{f}|\h{N-5},N-3,N-2,N-1,\dh \psi(N-2)|,\label{id-1}\\
  f|\h{N-4},N-2,&\dh \psi(N-2),\dt \psi(N-2)|\nonumber\\
  =&  -(b-\delta)^{N-2}\dh{f}|\h{N-4},N-2,N-1,\dt \psi(N-2)| \nonumber\\
  &+ (a-\delta)^{N-2}\dt{f}|\h{N-4},N-2,N-1,\dh \psi(N-2)|\nonumber\\
  =& (b-\delta)^{N-2}\dh{f}\Bigl[(a-\delta)^{N-2}\dt{s}
  +(a-\delta)^{N-1}\dt{g}+(a-\delta)^{N}\dt{f}-f \Bigr]\nonumber\\
  & -(a-\delta)^{N-2}\dt{f}\Bigl[(b-\delta)^{N-2}\dh{s}+
  (b-\delta)^{N-1}\dh{g}+(b-\delta)^{N}\dt{f}-f \Bigr],\label{id-2}
\\
 g|\h{N-4},N-2,&\dh \psi(N-2),\dt \psi(N-2)|\nonumber\\
 =&  -(b-\delta)^{N-2}\dh{f}|\h{N-4},N-2,N,\dt \psi(N-2)|\nonumber\\
  & + (a-\delta)^{N-2}\dt{f}|\h{N-4},N-2,N,\dh \psi(N-2)|,\label{id-3}
\\
(a-b)\dth{f}g=&\dh{f}[\dt{h}+(a-\delta)\dt{g}]-\dt{f}[\dh{h}
+(b-\delta)\dh{g}],\label{id-5}
 \\
  (a-b)(a-\delta)^{N-2}&(b-\delta)^{N-2}\dth{f}h
  = (a-\delta)^{N-2}\dt{f}|\h{N-3},N+1,\dh \psi(N-2)|\nonumber\\
  &\hspace{3cm}  -(b-\delta)^{N-2}\dh{f}|\h{N-3},N+1,\dt \psi(N-2)|.\label{id-6}
\end{align}
\end{subequations}

For the next identity we also need the following Lemma:
\begin{lem}\cite{Freeman-Nimmo-KP}
\label{L:iden}
\begin{equation}
\sum_{j=1}^{N}|\mathbf{a}_1,\cdots,\mathbf{a}_{j-1},\,
\mathbf{b}\mathbf{a}_{j},\, \mathbf{a}_{j+1},\cdots,
\mathbf{a}_{N}|=\biggl(\sum_{j=1}^{N}b_{j}\biggr)|\mathbf{a}_1,\cdots,
 \mathbf{a}_N|,
\label{iden-1}
\end{equation}
where $\mathbf{a}_j=(a_{1j},\cdots,a_{Nj})^T$ and
$\mathbf{b}=(b_1,\cdots,b_N)^T$ are $N$'th-order column vectors, and
$\mathbf{b}\mathbf{a}_j$ stands for $(b_{1}a_{1j}, \cdots, b_N
a_{Nj})^{T}$.
\end{lem}
Then, noting that
\begin{equation*}
(\delta-\omega k_j)^3-3\delta (\delta-\omega k_j)^2
+3\delta^2(\delta-\omega k_j)
\equiv (\delta- k_j)^3-3\delta (\delta- k_j)^2+3\delta^2(\delta- k_j)=\gamma_j,
\end{equation*}
we get, using Lemma \ref{L:iden} and the following identity
\begin{equation}
\Bigl[\Bigl(\sum^{N}_{j=1} \gamma_j\Bigr)\dt{f}\Bigr]\dh{f}=
\Bigl[\Bigl(\sum^{N}_{j=1} \gamma_j\Bigr)\dh{f}\Bigr]\dt{f},
\label{id-8}
\end{equation}
the explicit result
\begin{align}
~&(a-\delta)^{N-2}\dt{f}\Bigl[|\h{N-5},N-3,N-2,N-1,\dh \psi(N-2)|\nonumber\\
 & ~~~~~~~~~~~~~~~~~ - |\h{N-4},N-2,N,\dh \psi(N-2)|+|\h{N-3},N+1,\dh \psi(N-2)|\nonumber\\
 & ~~~~~~~~~~~~~~~~~ + (b-\delta)^{N+1}\dh{f}+g-(b-\delta)  f \nonumber\\
 & ~~~~~~~~~~~~~~~~~ -3\delta(b-\delta)^{N-2}( \dh{s}-\dh{h})-3\delta^2(b-\delta)^{N-2}\dh{g}\Bigr]\nonumber\\
 &-(b-\delta)^{N-2}\dh{f}\Bigl[|\h{N-5},N-3,N-2,N-1,\dt \psi(N-2)|\nonumber\\
 & ~~~~~~~~~~~~~~~~~~~ - |\h{N-4},N-2,N,\dt \psi(N-2)|+|\h{N-3},N+1,\dt \psi(N-2)|\nonumber\\
 & ~~~~~~~~~~~~~~~~~~~ + (a-\delta)^{N+1}\dt{f}+g-(a-\delta)  f \nonumber\\
 & ~~~~~~~~~~~~~~~~~~~ -3\delta(a-\delta)^{N-2}( \dt{s}-\dt{h})-3\delta^2(a-\delta)^{N-2}\dt{g}\Bigr]\nonumber\\
=&~0.
\label{id-7}
\end{align}

\subsection{Proof for bilinear equations \eqref{bil-DB-II}}\label{A:2}
With these Casoratian formulae derived in Appendix \ref{A:1}, we
can prove bilinear equations \eqref{bil-DB-II}.

\vskip 0.5cm\noindent \textbf{Proof for} \eqref{bil-DB-a2}:
We consider the down-tilde-shifted $\mathcal{B}^{\delta}_1$, i.e.,
\begin{equation}
{f}[\dt{h}+(a-\delta)\dt{g}]-{g}[\dt{g}+(a-\delta)\dt{f}]+\dt{f} {s}=0.
\end{equation}
Using \eqref{gh-mu}, \eqref{fg-mu} and \eqref{f-mu} we have
\begin{eqnarray*}
  &~& -(a-\delta)^{N-2}[{f}[\dt{h}+(a-\delta)\dt{g}]-{g}[\dt{g}+(a-\delta)\dt{f}]+\dt{f} {s}]\\
  &=&|\h{N-1}||\h{N-3},N,\dt{\psi}(N-2)| -|\h{N-2},N||\h{N-3},N-1,\dt{\psi}(N-2)|\\
  &~& +|\h{N-2},\dt{\psi}(N-2)| |\h{N-3},N-1,N|,
\end{eqnarray*}
which is zero in the light of  Lemma \ref{L:lap} by taking
$$\mathbf{B}=(\h{N-3}), ~~\mathbf{a}=\psi(N-2),~~\mathbf{b}=\psi(N-1),~~\mathbf{c}=\psi(N),~~\mathbf{d}=\dt{\psi}(N-2).$$
\eqref{bil-DB-b2} can be proved similarly.

\vskip 0.5cm\noindent\textbf{Proof for} \eqref{bil-DB-c2}:
We prove it in its down-tilde-hat-shifted version:
\begin{equation}
\dh{f}[\dt{g}+(a-\delta)\dt{f}]-\dt{f}[\dh{g}+(b-\delta)\dh{f}]-(a-b)f\dth{f}=0.
\label{id-4}
\end{equation}
This can be verified by using first \eqref{f-mu}, \eqref{fg-mu}, \eqref{f-th} and then Lemma \ref{L:lap} with
$\mathbf{B}=(\h{N-3})$, $\mathbf{a}=\psi(N-2)$, $\mathbf{b}=\psi(N-1)$, $\mathbf{c}=\dh\psi(N-2)$ and $\mathbf{d}=\dt{\psi}(N-2)$.

\vskip 0.5cm\noindent\textbf{Proof for} \eqref{bil-DB-d2}: We first
rewrite \eqref{bil-DB-d2} in the following form
\begin{equation}
f\Bigl[\dth{s}+(a+b+\delta)\dth{g}+(a^2+ab+b^2)\dth{f}\Bigr]
-g\Bigl[\,\dth{g}+(a+b+\delta)\dth{f}\Bigr]-(a^2+ab+b^2)\dt{f}\dh{f}+ \dth{f}h=0.
\label{bil-DB-d3}
\end{equation}
Then using \eqref{g-th} and \eqref{s-th} we have
\begin{eqnarray*}
  &~& (a-b)(a-\delta)^{N-2}(b-\delta)^{N-2}\times l.h.s. \eqref{bil-DB-d3}\\
  &=&f|\h{N-5},N-3,N-2,\dh{\psi}(N-2),\dt{\psi}(N-2)|\\
  &~& +(3\delta f-g)|\h{N-4},N-2,\dh{\psi}(N-2),\dt{\psi}(N-2)|\\
  &~& +(a-b)(a-\delta)^{N-2}(b-\delta)^{N-2}\dth{f}(3\delta^2 f-3\delta g+h)\\
  &~& +(a-\delta)^{N-2} \dt{f}[-(b+2\delta)f+g+b^3(b-\delta)^{N-2}\dh{f}]\\
  &~& -(b-\delta)^{N-2} \dh{f}[-(a+2\delta)f+g+a^3(a-\delta)^{N-2}\dt{f}].
\end{eqnarray*}
Next, we replace $f|\h{N-5},N-3,N-2,\dh{\psi}(N-2),\dt{\psi}(N-2)|$ by
\eqref{id-1}, $f|\h{N-4},N-2,\dh{\psi}(N-2),\dt{\psi}(N-2)|$ by
\eqref{id-2}, $g|\h{N-4},N-2,\dh{\psi}(N-2),\dt{\psi}(N-2)|$ by
\eqref{id-3}, $\dth{f} f$ by \eqref{id-4}, $\dth{f}g$ by \eqref{id-5}
and $\dth{f}h$ by \eqref{id-6}.  Then we find the remaining is nothing
but the l.h.s.of \eqref{id-7}, which is zero.  \qed

\subsection{Proof in the case of a generic $\Gamma$ matrix\label{G-proof}}
In fact, if $\psi$ satisfies \eqref{rela-I-a-1} and
\eqref{rela-I-a-2}, then we can get formulae \eqref{Formula-I} and we
only need a modification for the proof of the identity \eqref{id-7}.
In the proof we need the following
\begin{lem}
\label{L:id-III}\cite{Zhang-Hietarinta} (see also \cite{ZDJ-arxiv})
Suppose that $\Xi$ is an
  $N\times N$ matrix with column vector set $\{\Xi_j\}$;  $\Omega$ is an
  $N\times N$ operator matrix with column vector set $\{\Omega_j\}$ and
  each entry $\Omega_{j,s}$ being an operator. Then
  we have
\begin{equation}
\sum^N_{j=1} |\Omega_j * \Xi|
=\sum^N_{j=1}|(\Omega^T)_{j} * \Xi^T|,
\label{2.4}
\end{equation}
 where for any $N$'th-order column vectors $A_j$ and $B_j$ we define
\begin{equation*}
A_j \circ B_j=(A_{1,j}B_{1,j},~A_{2,j}B_{2,j},\cdots, A_{N,j}B_{N,j})^T
\end{equation*}
 and
\begin{equation*}
|A_j * \Xi|=|\Xi_1,\cdots,\Xi_{j-1},~A_j \circ\Xi_j,~\Xi_{j+1},\cdots, \Xi_{N}|
\end{equation*}
\end{lem}
In the light of the above Lemma taking $\Omega_{j,s}\equiv
(E^3)^3-3\delta (E^3)^2+3\delta^2E^3$ and $\Xi=\dt f$ or $\dh f$ we
have
\begin{equation}
(\mathrm{Tr}(\Gamma)\dt{f})\dh{f}=(\mathrm{Tr}(\Gamma)\dh{f})\dt{f}
\label{id-9}
\end{equation}
instead of \eqref{id-8}.  This equality together with formulae
\eqref{Formula-I} allows us to get the identity \eqref{id-7}.\qed
\end{appendix}

\end{document}